\documentclass[12pt]{article}
\usepackage{epsfig}
\begin{document}
\large

\title{Medium energy calorimetry at SND: techniques and performances 
on physics}
\author
{M.~N.~Achasov, V.~M.~Aulchenko, A.~V.~Berdyugin, A.~V.~Bozhenok, \\
A.~D.~Bukin, D.~A.~Bukin, S.~V.~Burdin, T.~V.~Dimova, \\ S.~I.~Dolinsky, 
A.~A.~Drozdetsky, V.~P.~Druzhinin, M.~S.~Dubrovin, \\ I.~A.~Gaponenko, 
V.~B.~Golubev, V.~N.~Ivanchenko, A.~A.~Korol, \\ S.~V.~Koshuba, 
G.~A.~Kukartsev, A.~P.~Lysenko, E.~V.~Pakhtusova, \\ E.~E.~Pyata, 
A.~A.~Salnikov, V.~V.~Shary, S.~I.~Serednyakov, \\ V.~A.~Sidorov,
Z.~K.~Silagadze \thanks {Corresponding author. Fax +7 3832 34 21 63,
e-mail silagadze@inp.nsk.su} ~, Yu.~V.~Usov, 
A.~A.~Valishev, \\ Yu.~S.~Velikzhanin, A.~S.~Zakharov \\
\vspace*{3mm} \\ Budker Institute of Nuclear Physics and \\
Novosibirsk State University, 630 090, Novosibirsk, Russia  }
\date{}

\maketitle

\begin{abstract}
SND (Spherical Neutral Detector) is a general purpose nonmagnetic detector,
successfully  operating  at Novosibirsk VEPP-2M collider  during last four 
years in the center of mass energy range from 400 MeV up to 1400 MeV. Its 
crucial part is a 3-layer, fine grained, spherical NaI(Tl) electromagnetic 
calorimeter consisted of 1632 individual counters. Here we present 
a detailed description  of the calorimeter and its  performance, including 
such topics as  calorimeter  design,  electronics,  calibration, energy and 
spatial resolutions,  particle identification.  The  calorimeter  performance
is illustrated  by examples from  the current  studies of different physical 
processes.
\end{abstract}

\section{Introduction}
Nowadays the Standard Model is triumphant -- the net result of high-energy 
physics experiments in the last two decades. The fact that for these stringent 
Standard Model checking experiments high energies are crucial is a natural
consequence of the W- and Z-boson high masses on one side, and of the 
asymptotic freedom of QCD on another. It is also true that the most part 
of the elementary particle physics community relates today further progress 
in the field to future experiments at even higher energies and is involved 
into the corresponding activities. Nevertheless many interesting and 
important things can be learned from experiments at low and medium energies, 
where QCD, one of the pillars of the Standard Model, becomes practically 
intractable because instead of the asymptotic freedom we have infrared 
slavery in this energy range and consequently a breakdown of the perturbation 
theory. 

For more than 25 years the VEPP-2M $e^+e^-$ collider in Novosibirsk operates
in the center-of-mass energy range $2E_0=0.36 \div 1.4$ GeV
\cite{vepp2}. Until now its maximum luminosity of
$L=3 \cdot 10^{30}~\mathrm{cm}^{-2}\mathrm{s}^{-1}$ at $E_0=510$ MeV
is a record in this energy range. During this period several
generations of detectors carried out experiments at VEPP-2M. Much of 
current data on low energy region particle properties \cite{pdg} 
were obtained in these experiments.

The SND detector, which calorimeter will be described here, is an advanced 
version of its predecessor -- the Neutral Detector (ND) \cite{ndnim,ndnm}, 
which successfully completed its five-year experimental program in 1987 
\cite{nd}. It also has some common features with the famous Crystal Ball 
detector \cite{CrBall}, but contrary to it has a multilayer structure
calorimeter, like Neutral Detector, and consequently more tools for 
$e/\pi$ and $\gamma/K_L$ separation.

The three-layer spherical electromagnetic calorimeter based on NaI(Tl)
crystals is the main part of the SND. It is optimized for investigation
of radiative decays of $\rho, \omega, \phi$ mesons, and other rare
processes at energy near 1 GeV with photons in the final state. A good 
energy and angular resolution for photons in the energy range from 
30 to 700 MeV is essential 
for background suppression in the $\pi^0$ and $\eta$ meson reconstructions 
and for detection of photons emitted in radiative transitions between 
quarkonia states.

\begin{figure}[htb]
  \begin{center}
\mbox{\epsfig{figure=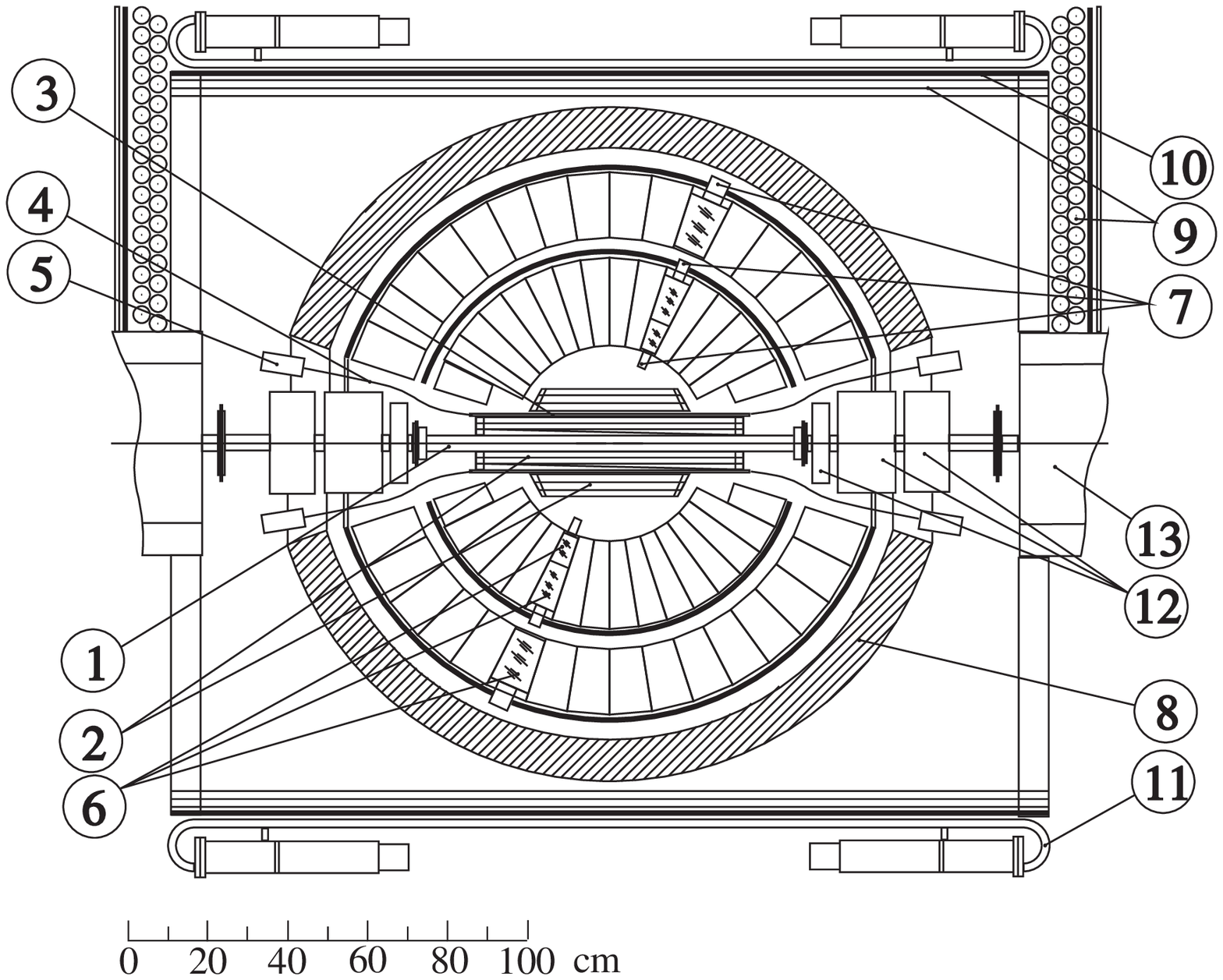
                               ,  height=12.5cm}}
   \end{center}
\caption {SND detector, section along the beams: (1) beam pipe,
         (2) drift chambers, (3) scintillation counter, (4) light guides,
         (5) PMTs, (6) NaI(Tl) crystals, (7) vacuum phototriodes,
         (8) iron absorber, (9) streamer tubes, (10) 1 cm iron plates,
         (11) scintillation counters, (12) and (13) elements of collider
         magnetic system. 
}
\label{sndt}
\end{figure}

\begin{figure}[htb]
  \begin{center}
\mbox{\epsfig{figure=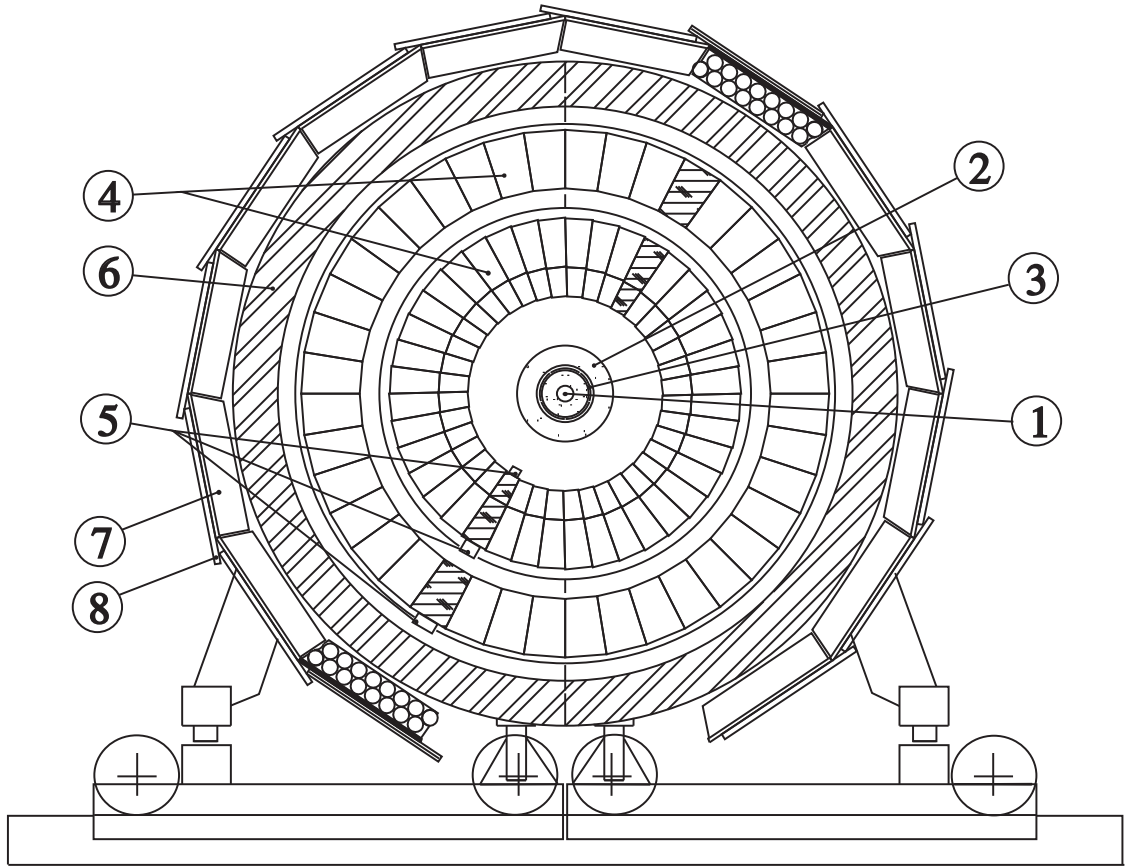
                               ,  height=11.5cm}}
   \end{center}
\caption {SND detector, section across the beams: (1) beam pipe,
         (2) drift chambers, (3) scintillation counter, (4) NaI(Tl)
         crystals, (5) vacuum phototriodes, (6) iron absorber,
         (7) streamer tubes, (8) scintillation system. 
}
\label{sndf}
\end{figure}

The general layout of the SND detector is shown in 
Figs.~\ref{sndt},~\ref{sndf}.
Electron and positron beams collide inside the beryllium beam pipe
with a diameter of 4 cm and thickness of 1 mm. The beam pipe is surrounded by
tracking system consisting of two drift chambers with a cylindrical
scintillation trigger counter between them.
The solid angle coverage of the tracking system is about $98\%$ of $4\pi$.
The electromagnetic calorimeter surrounds the tracking system. Outside the 
calorimeter a 12 cm thick iron absorber of the electromagnetic shower 
residuals is placed. It is surrounded by segmented muon system which provides  
muon identification and suppression of cosmic background. Each segment
consists of two layers of streamer tubes and a plastic scintillation
counter separated from the tubes by 1 cm iron plate.
The iron layer between the tubes and the counter reduces the
probability of their simultaneous firing by photons produced in
$e^+e^-$ collisions to less than $1\%$ for 700 MeV photons.

\section{Calorimeter layout}
\begin{figure}[htb]
  \begin{center}
\mbox{\epsfig{figure=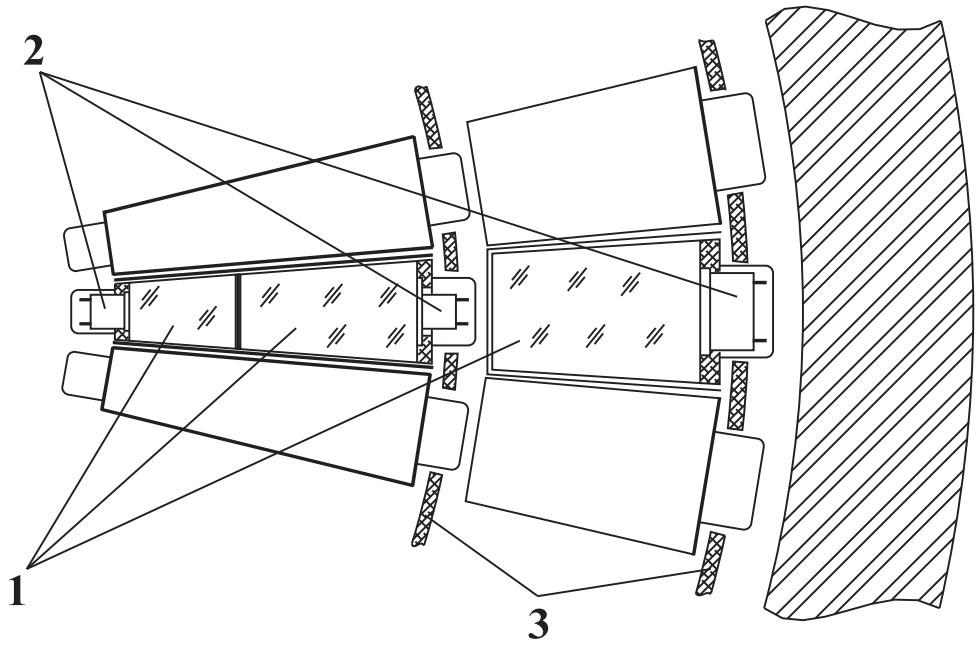
                               ,  height=8.5cm}}
   \end{center}
\caption {NaI(Tl) crystals layout inside the calorimeter: (1) NaI(Tl) 
crystals,
(2) vacuum phototriodes, (3) aluminum supporting hemispheres.
}
\label{cryst}
\end{figure}

Each layer of the calorimeter includes 520 to 560
crystals of eight different shapes. Most of the crystals have shapes of
truncated tetrahedral pyramids.  The total solid angle covered
by the calorimeter is equal to 0.9 of 4$\pi$.  The remaining space is
occupied by magnetic structure elements of the storage
ring, mainly by quadrupole lenses. Pairs of counters of the first two
layers with the thickness of 2.9$X_0$ and 4.8$X_0$ respectively ($X_0=2.6$~cm)
are sealed together in common containers made of thin (0.1 mm) aluminum foil
and fixed to an aluminum supporting hemisphere (Fig.~\ref{cryst}).
Behind it, the third layer of NaI(Tl) crystals, 5.7 $X_0$ thick, is placed.
In order to improve light collection efficiency and to separate one
crystal from another, each crystal is wrapped in an aluminized mylar. 
The gaps between adjacent crystals of one layer do not exceed 0.5~mm.
The total thickness of the calorimeter for particles originating from the 
center of the detector is 13.4 $X_0$ (34.7 cm) of NaI(Tl).
The total number of crystal in the calorimeter equals 1632 with 3.5 t
total mass.

\begin{figure}[htb]
  \begin{center}
\mbox{\epsfig{figure=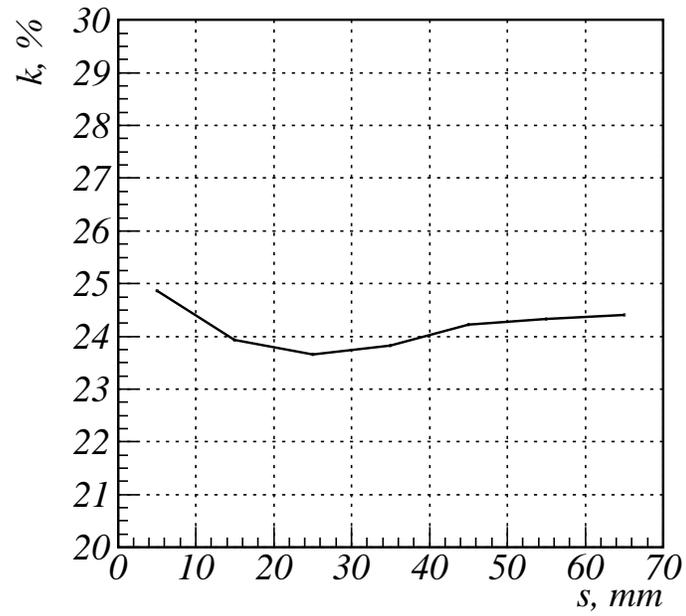
                               ,  height=8.5cm}}
   \end{center}
\caption {The typical behavior of the light collection efficiency along 
the first layer crystals.
}
\label{c1}
\end{figure}

\begin{figure}[htb]
  \begin{center}
\mbox{\epsfig{figure=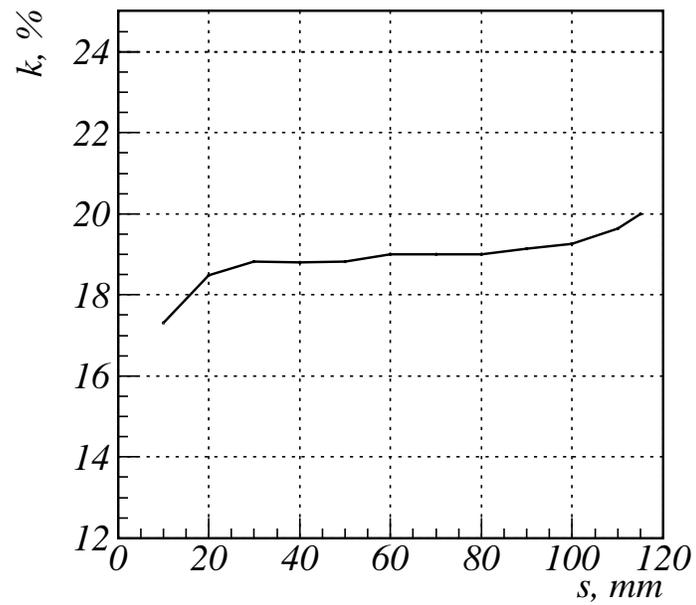
                               ,  height=8.5cm}}
   \end{center}
\caption {The typical behavior of the light collection efficiency along 
the second layer crystals.
}
\label{c2}
\end{figure}

During the crystals production their nonuniformity of the light collection 
efficiency along the crystal was controlled and only the crystals for which
the relative r.m.s. value of this nonuniformity was less than 3~\% were 
accepted. As number of crystals was large, special computer controlled 
measurement device was developed for this goal. A new ``quantum mechanical''
algorithm for automatic photopeak searches \cite{PhPeak} was used in the 
computer code for this device. The typical behavior of the light collection
efficiency along the crystal is shown in Fig.~\ref{c1} for the first layer 
crystals and in Fig.~\ref{c2} for the second layer ones. Crystals from the 
third layer are similar to the second layer crystals in this respect.
The light collection efficiency varies from $7\%$ to $25\%$ for crystals of
different calorimeter layers. The averaged efficiency for the first layer is
13\%, for the second layer -- 17\%, and for the third layer -- 18\%.

As photosensitive devices for the calorimeter counters the vacuum
phototriodes are used \cite{Beschastnov} with a photocathode diameter of 
17 mm in the first two layers and 42 mm in the third layer. The quantum 
efficiency of their photocathodes is about 15~\%, average gain is  10.
Crystals with lower light collection efficiency are equipped with higher gain
phototriodes and vice versa.

Spherical shape of the SND calorimeter provides relative uniformity of
its response over the whole solid angle. The range of polar angle coverage 
by the calorimeter is $18^\circ \leq \theta \leq 162^\circ$. The calorimeter 
is logically divided
into two parts:  ``large'' angles $36^\circ \leq \theta \leq 144^\circ$ and
``small'' angles --- the rest. The angular dimensions of crystals
are $\Delta\phi = \Delta\theta=9^\circ$ at large angles and
$\Delta\phi = 18^\circ, \Delta\theta=9^\circ$ at small angles.

The crystal widths approximately match the transverse size
of an electromagnetic shower in NaI(Tl). Two showers can be distinguished
if the angle between them is larger than $9^\circ$. If this angle
exceeds $18^\circ$ the energies of the showers can be measured with
the same accuracy as for an isolated shower. A high granularity of the
calorimeter is especially useful for the detection of multi-particle events.
For example, the detection efficiency for 7-photon events is usually close to 
15\% after all cuts needed to separate useful process from background.

For the trigger needs the calorimeter crystals are logically organized
into ``towers''. A tower consists of counters
located within the same $18^\circ$ interval in
polar and azimuthal directions
in all three layers. The number of counters in a tower is 12 at large
angles and 6 at small ones. 

\section{Electronics}
\begin{figure}[htb]
  \begin{center}
\mbox{\epsfig{figure=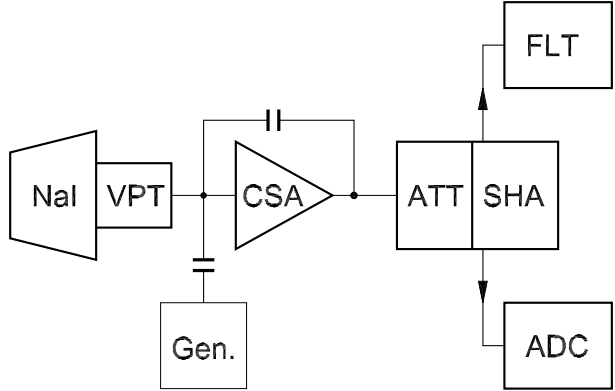
                               ,  height=8.5cm}}
   \end{center}
\caption {Electronics channel of the SND calorimeter: (NaI) NaI(Tl)
         scintillator, (VPT) vacuum phototriode, (CSA) charge-sensitive
         preamplifier, (ADC) analog to digital converter, (Gen) calibration
         generator, (SHA) shaping amplifier, (ATT) computer-controlled
         attenuator, (FLT) first-level trigger
}
\label{chan}
\end{figure}

Electronics of the calorimeter (Fig.~\ref{chan}) consists of
\begin{itemize}
\item
 the charge sensitive preamplifiers (CSA) with a conversion coefficient
 of~0.7~V/pC;
\item
 12-channel shaping amplifiers (SHA) with a remote controlled gain;
\item
 24-channel 12-bit analog to digital converters (ADC) with a maximum
 input signal $U_{\mathrm{max}}=2 \mathrm{V}$.
\end{itemize}
In order to equalize
contributions from
different counters into the total energy deposition signal and to obtain
equal energy thresholds for trigger signals
over the whole calorimeter, all shaping amplifiers
are equipped with computer controlled attenuators, allowing to adjust
SHA gain in steps of 1/255.

Each SHA serves one tower and also produces additional signals for 
the first level trigger,
the most important among them is the the total energy deposition ---
analog sum of all calorimeter channels in the tower. 
Other three logical signals correspond to energy depositions above 5 MeV
in separate calorimeter layers.

\begin{figure}[htb]
  \begin{center}
\mbox{\epsfig{figure=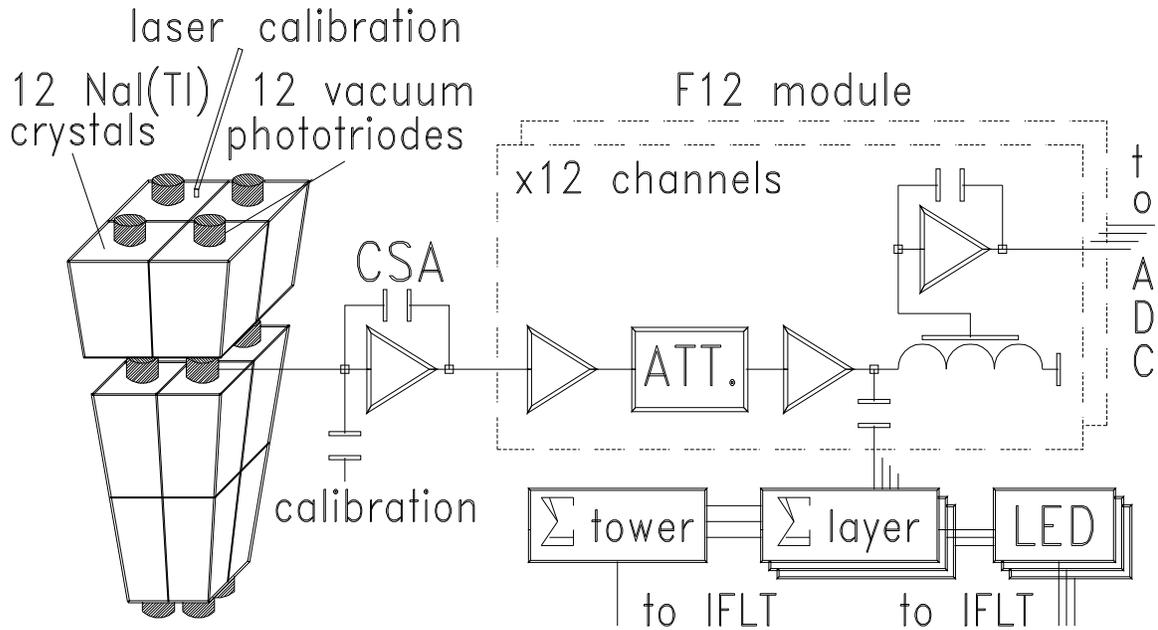
                               ,  height=8.5cm}}
   \end{center}
\caption{Calorimeter front-end electronics.}
\label{FendE}
\end{figure}

CSA and SHA constitute calorimeter front-end electronics (Fig.~\ref{FendE}).
The equivalent electronics noise of individual calorimeter channel lies 
within the range of $150\div350$ keV.

Each  calorimeter channel can be tested using
a precision computer-controlled calibration generator. The amplitude of its
signal can be set to any value from 0 to 1 V with a resolution of 1/4096.

The SND electronics is organized to provide necessary signals for 
the first level 
trigger (FLT). For example, the calorimeter shaping amplifiers are collected 
into 160 modules corresponding to 160 towers. From the FLT point of view 
the towers divide calorimeter into 20 sectors of $18^\circ$
in azimuthal direction and into 8 rings of $18^\circ$ in polar direction.

The following parameters are used for event identification in the 
calori\-meter FLT:
\begin{itemize}
\item total energy deposition in the calorimeter;
\item energy depositions in certain parts of the calorimeter;
\item number of clusters and their relative and absolute locations;
\item existence of a tower with energy depositions above a certain 
threshold in at least two calorimeter layers.
\end{itemize}

\begin{figure}[htb]
  \begin{center}
\mbox{\epsfig{figure=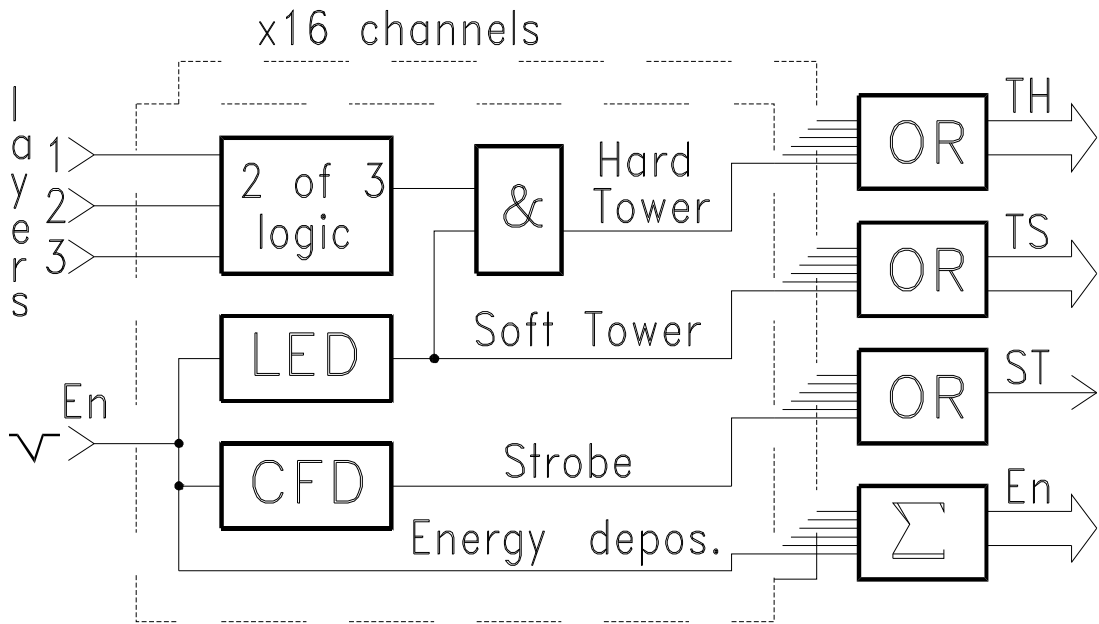
                               ,  height=8.5cm}}
   \end{center}
\caption{IFLT module layout.}
\label{IFLT}
\end{figure}

Signals from 16 towers, forming two calorimeter sectors, are collected 
by  IFLT modules (Fig.~\ref{IFLT}). For each tower two logical signals are 
produced: a ``soft tower'' (TS) with a total energy deposition higher than 
25 MeV, and a ``hard tower'' -- the same as a TS but with energy depositions
in two of three layers larger than 5 MeV.

\begin{figure}[htb]
  \begin{center}
\mbox{\epsfig{figure=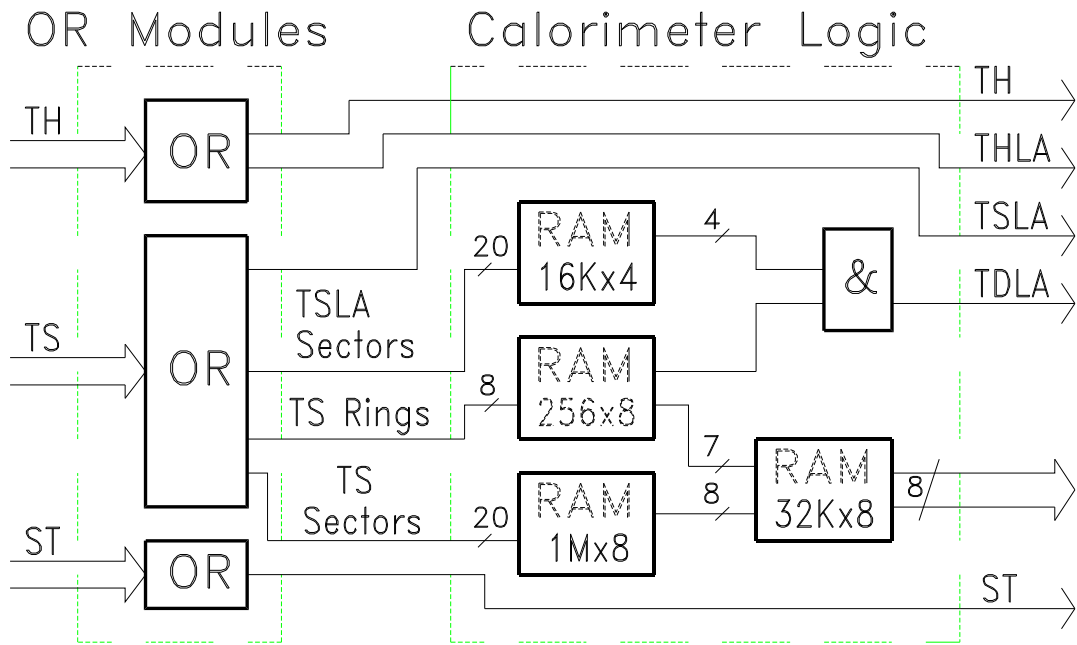
                               ,  height=8.5cm}}
   \end{center}
\caption{Calorimeter Logic module.}
\label{calog}
\end{figure}

Fig.~\ref{calog} shows the organization of the Calorimeter Logic module. Two
groups of 20 signals each from sectors at different angles, produced by the 
summation of tower signals in the polar direction, and 8 signals from rings 
-- the sums of towers in the azimuthal direction -- form the set of input
signals for the Calorimeter Logic module.

The ring signals and the partial sector signals from towers at large angles 
(TSLA) produce the signal ``two towers at large angle'' (TDLA). The 
sector signals (TS) go to the address bus of 1MB static RAM, where the
look-up table is stored. The information from rings contained in another
32 kB RAM is added. So, according to the memory contents, the components
of the FLT defining the number of towers and their relative positions are
produced. FLT components can be easily changed by RAM reprogramming. 

\section{Calibration}
\begin{figure}[htb]
  \begin{center}
\mbox{\epsfig{figure=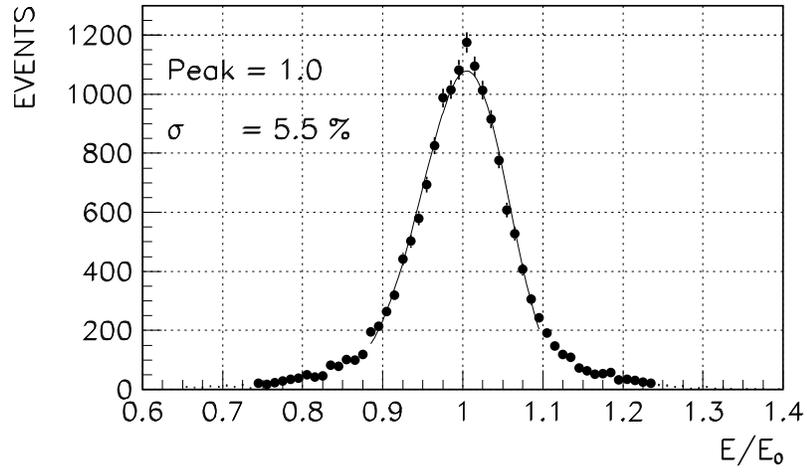
                               ,  height=8.5cm}}
   \end{center}
\caption{Energy spectra for photons with energy $E_0 = 500$ MeV after cosmic
         calibration. $E$ is a measured energy.}
\label{ggcc}
\end{figure}

Calorimeter is calibrated using cosmic muons \cite{ccc} and
$e^+e^- \rightarrow e^+e^-$ Bhabha events \cite{ecc}.
A fast preliminary calibration based on cosmic muons gives the constants for
the energy deposition calculations in the calorimeter crystals. These constants
are used  to set computer controlled attenuators at such values for each 
calorimeter channel that equal responses of all crystals are obtained. This
provides uniform first-level trigger energy threshold over the whole 
calorimeter. They also 
represent seed values for the more precise calibration procedure,
using $e^+e^- \rightarrow e^+e^-$ events. The cosmic calibration procedure
is based on the comparison of the experimental and simulated
energy depositions  in the calorimeter crystals for cosmic muons.
The detailed description of the method is given in  \cite{ccc}. The statistical
accuracy of $1\%$ in the calibration coefficients was achieved for about 1.5
million events collected in special data taking runs with total duration 
of 4.5 hours. After cosmic calibration the peak positions in the measured 
energy spectra for photons and electrons agree at a level of about $1\%$ 
with the actual particle energies (Fig.~\ref{ggcc}).

The cosmic calibration procedure was performed weekly, between the
experimental runs.For a one-week period between consecutive
calibrations the coefficients stability was better than $1.5\%$. Changes in
the coefficients were also monitored daily, using the calibration generator.
They could either drift slowly due to electronics gain changes or show
larger leaps when broken electronics modules are replaced.

Crystal-phototriode-preamplifier channel short term stability was also 
previously studied
in series of measurements with a calorimeter segment by using laser-based 
calibration system with a fiber optics \cite{SNDCal}. Light pulses from 
a nitrogen laser after distribution system was driven to both counter crystals
and reference photodiodes which measure the pulse amplitude. Instability 
shown by this laser calibration was about 0.4\% \cite{SNDCal}.

To achieve highest possible energy resolution, the precise OFF-LINE
calibration procedure, based on the Bhabha 
$e^+e^- \rightarrow e^+e^-$ events analysis, was implemented.
The calibration coefficients are obtained by the minimization
of the r.m.s. of the total energy deposition  spectrum for the electrons 
with fixed energy. The detailed description of the procedure is given in 
\cite{ecc}. To
obtain the statistical accuracy of about $2\%$ the sample of
at least 150 electrons per crystal is needed. SND acquires such a sample
daily when VEPP-2M operates in the center-of-mass energy $2E_0 \sim 1$ GeV.
The average difference in calibration coefficients obtained using cosmic
and $e^+e^-$ calibration procedures is about $4\%$, while the $e^+e^-$
calibration improves the energy resolution by about $10\%$. For example, 
the energy resolution for 500 MeV photons improves from $5.5\%$ to $5.0\%$.

\section{Energy and spatial resolutions}
\begin{figure}[htb]
  \begin{center}
\mbox{\epsfig{figure=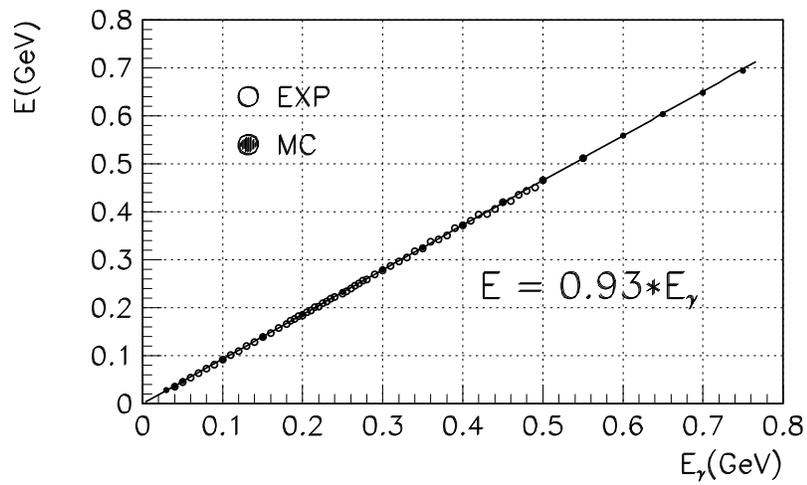
                               ,  height=8.5cm}}
   \end{center}
\caption{Dependence of the most probable energy deposition on the photon
         energy.}
\label{gali}
\end{figure}

The calorimeter energy resolution is determined mainly by the fluctuations
of the energy losses in the
passive material before and inside the calorimeter and leakage of
shower energy through the calorimeter.
The most probable
value of the energy deposition for photons in the calorimeter is about $93\%$
of their energy (Fig.~\ref{gali}). The estimated energy losses in passive
material for 500 MeV photons obtained by simulation are
listed in Table~\ref{tab51}.
\begin{table}[hb]
\small
\begin{center}
\begin{tabular}[t]{|c|c|c|c|}
\hline
 element & thickness $(X_0)$ & $E(\%)$ & $\sigma(\%)$ \\ \hline
 passive material in front of the calorimeter &
           0.17            & 0.16    & 0.05         \\ \hline
 passive material between 2nd and 3d layer &
           0.17            & 0.84    & 0.96         \\ \hline
 longitudinal leakage &
           ---             & 3.5     & 1.2 \\ \hline
 transverse leakage &
           ---             & 2.8     & 1.1          \\ \hline
 containers     &
           ---             & 0.26    & 0.05         \\ \hline
 NaI(Tl) &  13.4           & 93      & 3.3          \\ \hline
\end{tabular}
\caption{The most probable energy deposition  $E(\%)$ at calorimeter
         elements for 500 MeV photons.}
\label{tab51}
\end{center}
\end{table}

In order to compensate for the shower energy losses in passive material
and improve energy resolution the photon energy is calculated as:
\begin{equation}
 E = \alpha_1 \cdot E_1 + \alpha_2 \cdot E_2 + \alpha_3 \cdot E_3, \label{eq:x}
\end{equation}
where $E_1$ is energy deposition in the first and second layers of
the central tower of the shower, $E_2$ is energy deposition in the first
two layers outside the central tower, $E_3$ is energy deposition in the third
layer, $\alpha_i$ are energy dependent coefficients. Here the tower means the
three counters of the 1, 2 and 3 layers with the same $\theta$ and $\phi$
coordinates and the central tower corresponds to the shower center of gravity.

The $\alpha_i$ coefficients were determined from simulation of
photons with energies from 50 to 700 MeV. For each photon
energy the following function:
 \begin{equation}
 M = \sum\limits_{k}(E^* - E_k )^2,
 \end{equation}
was minimized with regard to $\alpha_i$.
Here $E^*$ is a known photon energy,
$E_k$ is the energy calculated using expression (\ref{eq:x}). The energy
dependences of $\alpha_i$ were approximated by the smooth curves. The
approximation was done separately for showers starting
in different layers.

\begin{figure}[htb]
  \begin{center}
\mbox{\epsfig{figure=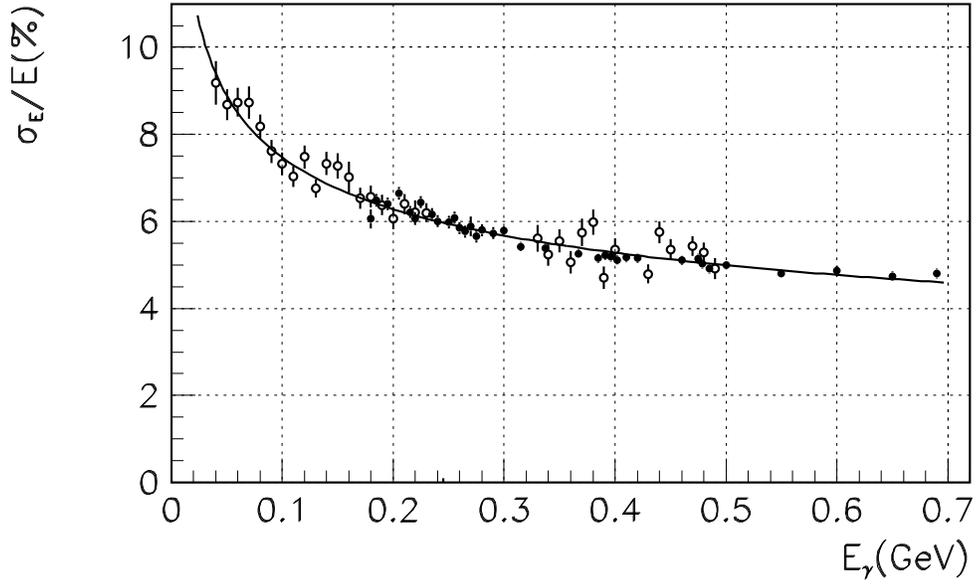
                               ,  height=8.5cm}}
   \end{center}
\caption{Dependence of the calorimeter energy resolution on the photon energy,
         $E$ -- photon energy, $\sigma_E/E$ -- energy resolution obtained
         using $e^+e^- \rightarrow \gamma \gamma$ (dots) and
         $e^+e^- \rightarrow e^+e^- \gamma$ (circles) reactions.
         }
\label{resge}
\end{figure}

The apparatus effects, such as nonuniformity of light collection
efficiency over the crystal
volume and electronics instability also affect energy resolution:
\begin{equation}
 \sigma_E/E(\%) = \sigma_1(E) \oplus \sigma_2(E) \oplus \sigma_3(E),
\end{equation}
where $\sigma_1(E)$ is the energy resolution obtained using Monte Carlo
 simulation without effects mentioned above,
$\sigma_2(E)$ is electronics instability and calibration accuracy
contribution, $\sigma_3(E)$ is the contribution of nonuniformity of light
collection over the crystal volume. For example, for 500 MeV
photons $\sigma_E/E = 5\%$, $\sigma_1(E)=3\%$, $\sigma_2(E)=1.2\%$ and
$\sigma_3(E)=3.8\%$. The dependence of the calorimeter energy resolution on
photon energy (Fig.~\ref{resge}) was fitted as:
\begin{equation}
 \sigma_E/E(\%) = {4.2\% \over \sqrt[4]{E(\mathrm{GeV})}}.
\end{equation}
 
\begin{figure}[htb]
  \begin{center}
\mbox{\epsfig{figure=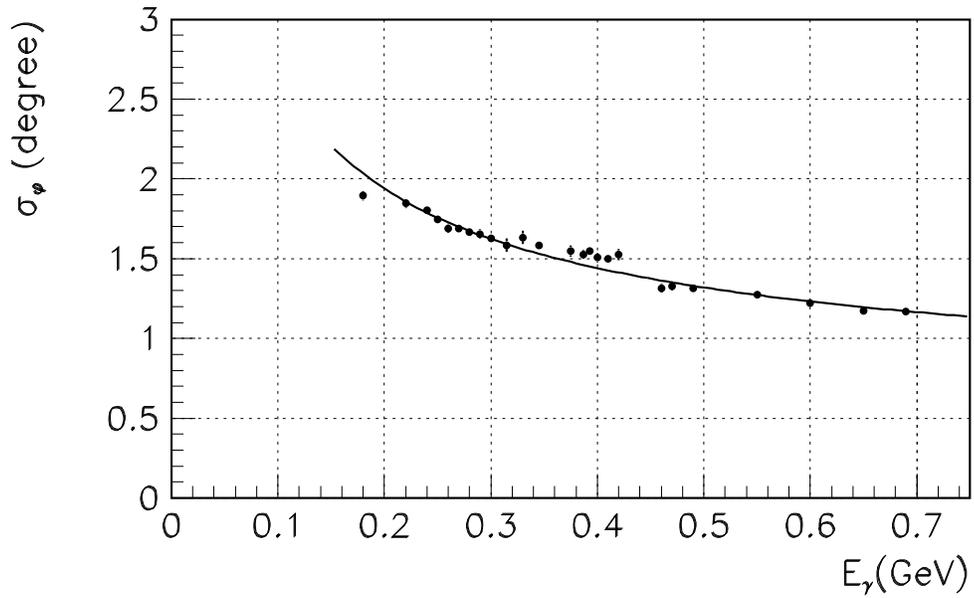
                               ,  height=8.5cm}}
   \end{center}
\caption{Dependence of the angular resolution on the photon energy,
        $E$ -- photon energy.
         }
\label{resan}
\end{figure}

The calorimeter angular resolution for photons is also an important parameter.
The distribution function of energy deposition in SND calorimeter outside
the cone with the angle $\theta$ around the shower direction was obtained
using Monte-Carlo simulation:
\begin{equation}
  E(\theta) = \alpha \cdot exp(-~\sqrt[]{\theta/\beta}),
\end{equation}
where the $\alpha$ and $\beta$ parameters turned out to be practically
independent of the photon energy over the interval $50 \div 700$ MeV.
The method  of estimation of photon angles based on this dependence was
introduced in \cite{ivbe}. The dependence of the angular resolution on
the photon
energy is shown on  Fig.~\ref{resan} and can be approximated as:
\begin{equation}
 \sigma_\phi = {0.82^\circ \over \sqrt[]{E(\mathrm{GeV})}} \oplus 0.63^\circ.
\end{equation}

\section{Particle identification}

\begin{figure}[htb]
  \begin{center}
\mbox{\epsfig{figure=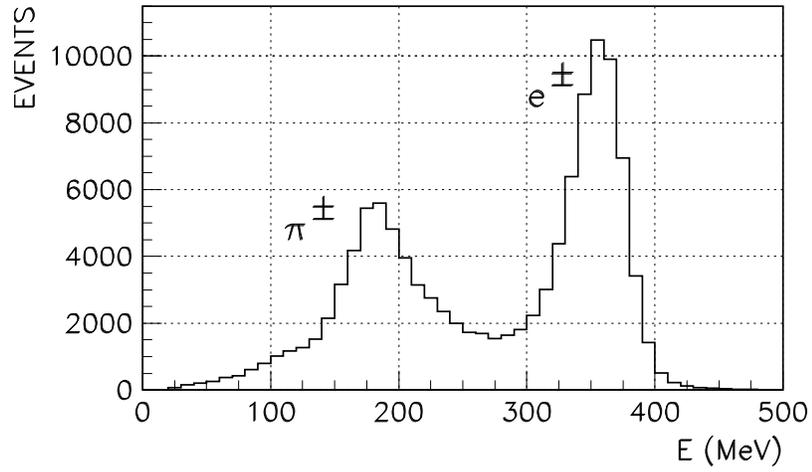
                               ,  height=8.5cm}}
   \end{center}
\caption{Energy deposition spectra for 385 MeV $e^\pm$ and $\pi^\pm$}
\label{sep1}
\end{figure}

\begin{figure}[htb]
  \begin{center}
\mbox{\epsfig{figure=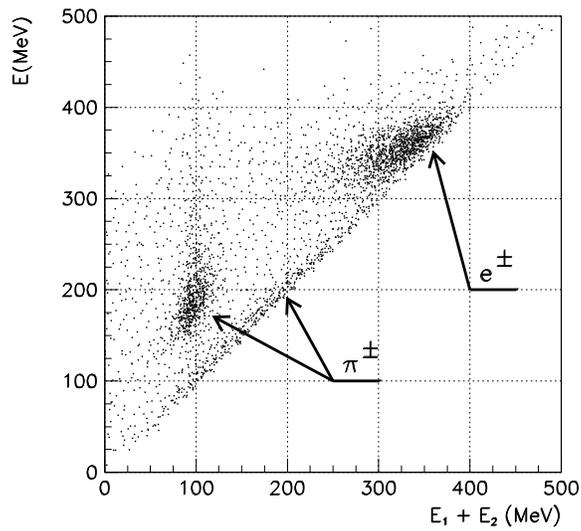
                               ,  height=8.5cm}}
   \end{center}
\caption{Energy deposition in calorimeter first two layers vs total
         energy deposition for 385 MeV $e^\pm$ and $\pi^\pm$. $E$ -- total
         energy deposition, $E_1+E_2$ -- energy deposition in the first
         two layers}
\label{sep2}
\end{figure}

The discrimination between electromagnetic and hadronic
showers in the calorimeter is often based on the total energy
deposition  (Fig.~\ref{sep1}).
Multilayer structure of the SND calorimeter provides additional
means for identification based on
the longitudinal energy deposition profiles.
energy deposition distributions over layers for $e^\pm$ and
$\pi^\pm$ are shown in Fig.~\ref{sep2}.
Two areas of pion concentration on this scatter plot
correspond to the nuclear interactions of pions and pure ionization losses
in the first two layers of NaI(Tl).
Utilizing differences in energy depositions for electrons
and pions, the special discrimination parameter was worked out. In the
$\rho(770)$ energy region it provides $99\%$ selection efficiency for
$e^+e^- \rightarrow \pi^+\pi^-$, while keeping contamination by
$e^+e^- \rightarrow e^+e^-$ events $\sim 1\%$.

The distribution of energy depositions in the calorimeter layers is also
used for $K_L$
identification in $\phi \rightarrow K_S K_L$ decay. Such an identification
provides the suppression of background in searches of rare $K_S$ decays,
for example $K_S \rightarrow 3\pi^0$ \cite{Ks3pi}.

The $e/\pi$ and $\gamma/K_L$ separation parameters, based on the differences 
in the
energy deposition profiles in transverse direction, were also constructed.
Their detailed description is given in \cite{ivbo}.

\section{The calorimeter performance}
\begin{figure}[htb]
  \begin{center}
\mbox{\epsfig{figure=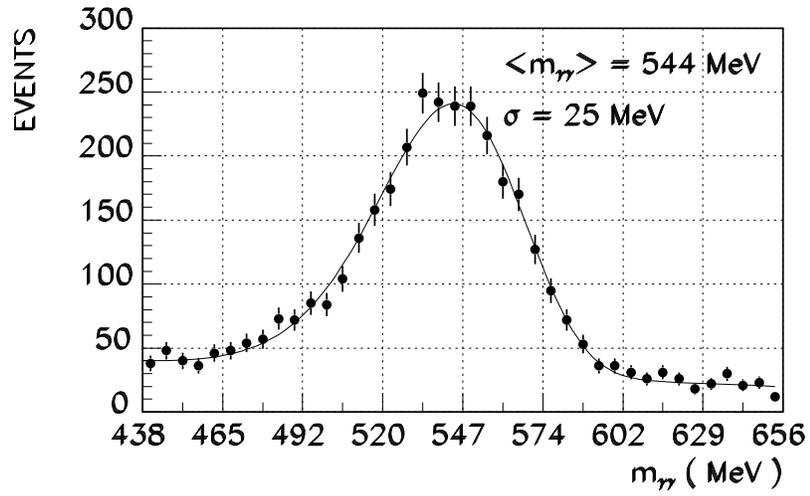
                               ,  height=8.5cm}}
   \end{center}
\caption{Two photon invariant mass distribution in experimental
         $\phi \rightarrow \eta \gamma$ events.}
\label{etag}
\end{figure}

\begin{figure}[htb]
  \begin{center}
\mbox{\epsfig{figure=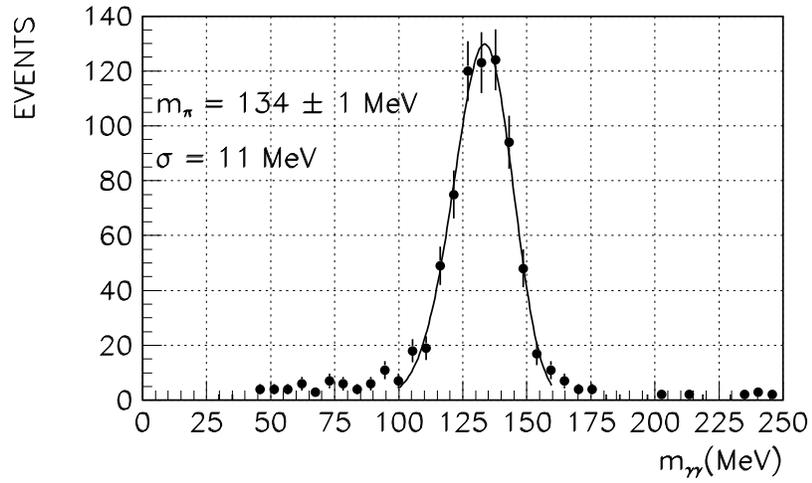
                               ,  height=8.5cm}}
   \end{center}
\caption{Two photon invariant mass distribution in experimental
         $\phi \rightarrow \pi^+ \pi^- \pi^0$ events.
         }
\label{pi3}
\end{figure}

The calorimeter energy and angular resolution determine the accuracy of
invariant mass measurements. The two-photon invariant mass distributions
in $\phi \rightarrow \eta \gamma$ and
$\phi \rightarrow \pi^+\pi^-\pi^0$ events, depicted in
Figs.~\ref{etag},~\ref{pi3}, show clear peaks at $\pi^0$ and $\eta$
mesons masses. Invariant mass resolution is about 11 MeV for $\pi^0$
and 25 MeV for $\eta$.

\begin{figure}[htb]
  \begin{center}
\mbox{\epsfig{figure=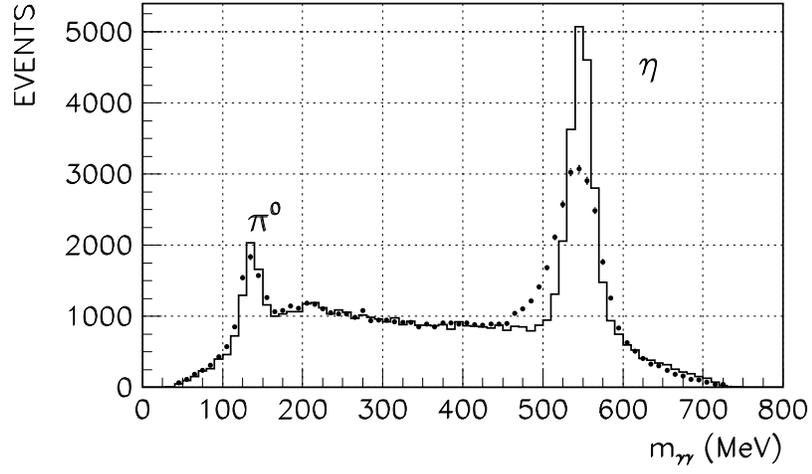
                               ,  height=8.5cm}}
   \end{center}
\caption{Two photon invariant mass distribution in the
         $e^+e^- \rightarrow \gamma \gamma \gamma$ events before (dots) and
         after (line) kinematic fitting.
         }
\label{ggg}
\end{figure}

For the physical analysis of the most processes the kinematic fitting
\cite{kin1} is performed. As a result the angular and energy accuracy
improve. For example, in the analysis of
$\phi \rightarrow \eta \gamma, \pi^0 \gamma$ decays the fitting
increases the two photons invariant mass resolution by 1.5 times
(Fig.~\ref{ggg}).

\begin{figure}[htb]
  \begin{center}
\mbox{\epsfig{figure=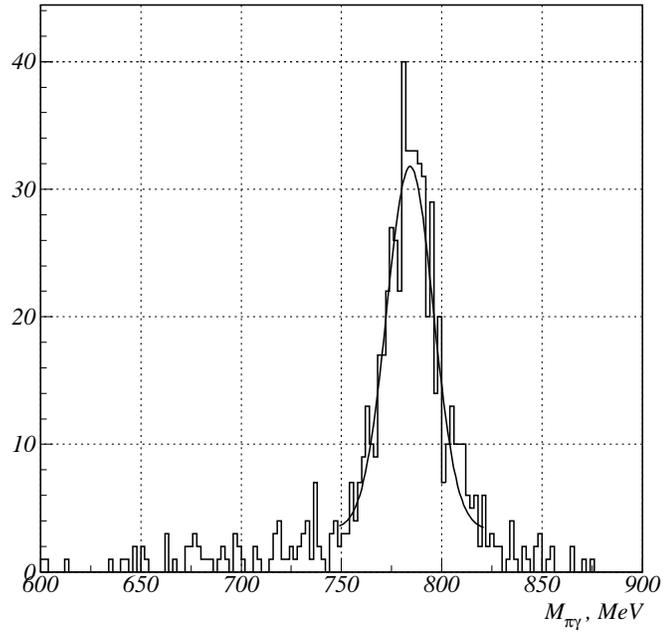
                               ,  height=8.5cm}}
   \end{center}
\caption{$M_{\pi^0\gamma}$ (invariant mass of $\pi^0\gamma$ nearest to
$\omega$) distribution for the process 
$e^+e^-\to\omega\pi^0\to\pi^0\pi^0\gamma$}
\label{wpi}
\end{figure}

As a last example of the calorimeter performance, Fig.~\ref{wpi} shows 
$\omega$-meson peak in the reaction $e^+e^-\to\omega\pi^0\to\pi^0\pi^0\gamma$.

\section{Conclusion}
SND detector operates at VEPP-2M since 1995. The total integrated luminosity
$\sim~25$~$pb^{-1}$ in the center-of-mass energy range $2E_0 = 0.4 \div 1.4$
GeV was collected. Electric dipole radiative decays
$\phi \rightarrow \pi^0 \pi^0 \gamma$ \cite{f0g}, $\eta \pi^0 \gamma$
\cite{a0g} and OZI and G-parity suppressed decay
$\phi \rightarrow \omega \pi^0$ \cite{ompi0} were observed for the first time.
The $\phi \rightarrow \eta^\prime \gamma$ \cite{etg} decay existence was
confirmed. Other interesting physical results were also obtained. 

\begin{figure}[htb]
  \begin{center}
\mbox{\epsfig{figure=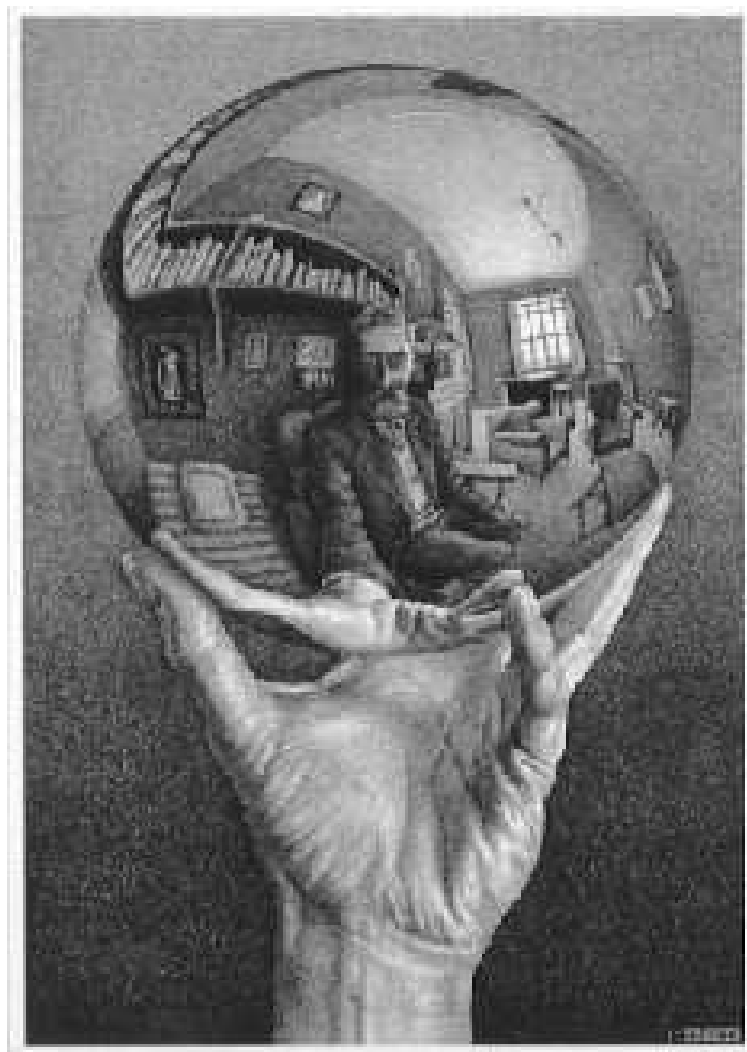
                               ,  height=14cm}}
   \end{center}
\caption{Artist's view on the SND calorimeter.}
\label{Escher}
\end{figure}

The SND calorimeter performance was crucial for all these results.
In the last Fig.~\ref{Escher} we present an artist's view on the 
SND calorimeter and provide citation from ``M.C. Escher - The Graphic Work''
\cite{Escher}, which gives, in our opinion, a concise impression
to what the calorimeter is used for and also exhibits a real driving force in
all high and medium energy physics experiments, and not only there.

``A reflecting globe rests in the artist's hand. In this mirror he can have 
a much more complete view of his surroundings than by direct observation, 
for nearly the whole of the area around him - four walls, the floor and 
ceiling of his room - are compressed, albeit distorted, within this little 
disc. His head, or to be more precise, the point between his eyes, comes in 
the absolute center. The ego is the unshakable core of his world.''

\newpage

\clearpage

\newpage

\end{document}